# Application of Sequence Embedding in Protein Sequence-Based Predictions

Nabil Ibtehaz[1], Daisuke Kihara[1,2,*]

1 Department of Computer Science, Purdue University, West Lafayette, IN, United States

2 Department of Biological Sciences, Purdue University, West Lafayette, IN, United States

Corresponding Author, dkihara@purdue.edu

**Abstract**

In sequence-based predictions, conventionally an input sequence is represented by a multiple sequence alignment (MSA) or a representation derived from MSA, such as a position-specific scoring matrix. Recently, inspired by the development in natural language processing, several applications of sequence embedding have been observed. Here, we review different approaches of protein sequence embeddings and their applications including protein contact prediction, secondary structure, prediction, and function prediction.

## 1 Introduction

Proteins are the fundamental building blocks of life, driving the entire network of interconnected and intercorrelated functional mechanisms in an organism [1]. Proteins are involved in almost every cellular function, including signaling pathway, DNA repair, glucose transmembrane transport, catalytic activity, and transporter activity. The tertiary structures of proteins and probably also interactions protein interactions are encoded in the sequence of amino acids. Thus, protein sequences are often termed as the *language of life* [2].

Analyzing protein sequences and inferring various functional and structural information has been one of the major goals and long-standing themes of Bioinformatics. Next-generation sequencing technologies have led to an exponential increase in the size of protein databases, nearly doubling almost in every two years [3]. However, labeling them with valid and meaningful annotations requires an extensive amount of effort, expertise, experiments, and expense. As a result, we have new proteins in orders of magnitudes more than it is viable to annotate manually. This discrepancy becomes more apparent when we observe that the sparsely annotated TrEMBL database contains 219 million sequences, whereas the manually curated SwissProt database contains only 565 thousand proteins [3]. Thus, the so-called sequence-structure gap [4] is increasingly growing.

Bioinformatics researchers have devoted decades in developing various computational prediction methods of protein structures, features, and annotations from amino acid sequence information. In conventional prediction methods, typically, sequence information of a target protein is provided in forms including a single protein sequence, position-specific scoring matrix (PSSM) [5], Hidden Markov Model (HMM) [6], and k-grams. Often, physicochemical properties of amino acids, such as hydrophobicity, charge, and size information are also used, instead of, or in addition to, the amino acid sequence itself [7].

In recent years, the field of Natural Language Processing (NLP) has observed a radical paradigm shift, by embracing pre-trained language models [9, 10]. The trend has been to train a language model on a large corpus of unlabeled text data in an unsupervised or semi-supervised fashion [11], which enables the models to learn patterns and structures of the language. This pre-training provides us with a general knowledge of the language in form of embeddings, which are found to be effective in solving various downstream tasks, occasionally with the aid of some task-specific finetuning. These pre-trained embedding approaches significantly improved upon the earlier supervised methods trained on task-specific smaller datasets [9].

Deriving motivations from the success of word embeddings and pretrained language models in NLP, gradually they are gaining popularity in protein sequence analysis. Several language models have been adopted and applied for proteins, for example, ProtVec [12], SeqVec [2], and ProtBERT [13].

In this chapter, we explore the application of sequence embedding in protein sequence-based predictions. We briefly explain some notable language models and how they are used in bioinformatics, facilitated by access to large-scale protein databases. We present the effectiveness of learned sequence embeddings in solving various problems on diverse topics.

## 2   A brief overview of language models and embeddings in Natural Language Processing

First, we briefly present a few prominent language models and corresponding embedding generation schemes used in Natural Language Processing (NLP). For a more elaborate explanation, readers are encouraged to go through broader surveys [9, 10].

Distributed representations [14] and Neural Network-based language models [15] have a long history of gradual progressive development. The first significant breakthrough came from the works of Mikolov et al. [16, 17]. They proposed a novel word embedding named word2vec, which represents the words as dense vectors in a relatively low dimensional space. Fig. 1(a) presents a simplified representation of word2vec, where we show how the input-output can be represented and computed in the skip-gram model. The embeddings are learned from a shallow neural network, by analyzing the neighboring words using Log-Linear models such as Continuous Bag-of-Words, which tries to predict the word based on the context, or Skip-gram, which attempts to predict the context based on the word. Both these two approaches are based on the fact that not only the semantic but also the syntactic meaning of a word can be estimated by its neighboring words [17]. In this word embedding, similar words are projected nearby in the vector space, thus providing a means of comparing the words both syntactically and semantically, which turns out valuable in various downstream NLP tasks.

Despite the effectiveness of mapping similar and dissimilar words, one major problem the word2vec model faces is that the generated embeddings are context independent. In natural language, the same word can have multiple meanings based on the context it is used. Word2vec discards this contextual and positional dependency of words. Various Long-Short Term Memory (LSTM) based models, for example, ASGD Weight-Dropped LSTM (AWD-LSTM) [18] (Fig. 1b), multiplicative LSTM (mLSTM) [19] (Fig. 1c), and more specifically, ELMo (Embeddings from Language Models) [20] (Fig. 1d) targets to solve this issue, by treating the word embeddings not merely as a function of the words, rather as a function of the entire sentence. LSTM, being a recurrent neural network, keeps track of the order of the words unlike the feed-forward networks used in word2vec. As a result, it manages to learn sentences not just merely a collection of words. AWD-LSTM investigates several strategies for regularizing and optimizing LSTM models, incorporating various levels of dropout at inputs, hidden layers, weights, embeddings, and outputs. Furthermore, the model was trained with a modified averaged stochastic gradient method, and it outperformed other models on word-level perplexities, a standard language model task, where a model's ability to compute the probability of unseen test sentences is evaluated. on two datasets, Penn Treebank and WikiText-2. On the other hand, mLSTM combines the standard LSTM with multiplicative RNN architecture. This provides mLSTM with the ability to have different recurrent transition functions for different inputs and this increased expressivity. This enables mLSTM to consistently outperform vanilla LSTM on character-level language modeling task. Among the various LSTM based models, ELMo produces the most superior contextualized word representation, managing to capture the syntax and semantics of the words across various linguistic contexts. ELMo achieves this through a semi-supervised setup, by pretraining a Bidirectional Language Model (biLM) on large-scale datasets, later incorporating that with diverse NLP architectures. The biLM aims to model the probability of a token $t_k$, in a sequence ($t_1,t_2,t_3,...,t_N$) by considering both the history and the future contexts.

$$p(t_1, t_2, t_3, \ldots, t_N) = \begin{cases} \prod_{k=1}^{N} p(t_k|t_1, t_2, \ldots, t_{k-1}) & \text{, forward LM} \\ \prod_{k=1}^{N} p(t_k|t_{k+1}, t_{k+2}, \ldots, t_N) & \text{, backward LM} \end{cases} \quad (1)$$

In the process of learning these probabilities, each of the L layers of the BLSTMs outputs a context-dependent representation for a token $t_k$ in both directions $\vec{h}_{k,j}^{LM}, \overleftarrow{h}_{k,j}^{LM}$, where $j = 1,2,..., L$. Combining these with a context-independent representation $x^{LM}_k$, ELMo embeddings are computed. By incorporating the context of the words, the issue of homonyms is thus solved by ELMo, as well as capturing high-level concepts from context. AWD-LSTM significantly improves over the traditional LSTM models. With the efficient utilization of the regularizations, it managed to reduce perplexity by 20%, despite requiring 1/3 of the parameters. Although mLSTM managed to decrease perplexity by 10% over vanilla LSTM, it was outperformed by the AWD-LSTM model, due to it being a character-level architecture. On the other hand, as a result of using bidirectional LSTM layers, ELMo appears to be the most capable LSTM based model, going beyond the state of the art performance in 6 tasks, including question answering, named entity recognition, and sentiment analysis.

The LSTM based models AWD-LSTM, mLSTM, and ELMo have been illustrated in Figures 1b, 1c, and 1d respectively. AWD-LSTM is defined by the repertoire of dropouts it employs, as demonstrated in Fig. 1b. mLSTM (Fig. 1c) on the other hand, disregards the typical hidden state in LSTM, rather maps the hidden state *h(t-1)* to an input dependent one *m(t)* by suitable weighted multiplication with the input *x(t)*. The foundational component of ELMo (Fig. 1d) is the BLSTM base, which is accompanied by embeddings generated from character convolutions and suitable softmax top layers that address a particular downstream task.

ELMo improves over word2vec based embedding, but the next breakthrough came from Transformer based architectures, most notably BERT (Bidirectional Encoder Representations from Transformers) [11]. Transformer is a novel network architecture based on attention [21], instead of traditional convolutional or recurrent operations. Discarding recurrent operations makes the transformer model parallelizable, which reduces the training time and improves the performance simultaneously. Transformer is composed of an encoder and a decoder, which themselves contain stacks of layers involving multi-head attention and position-wise feedforward networks, accompanied by residual connections and layer normalizations. A simplified schematic diagram of Transformer has been presented in Figure 1(e). Transformers have an encoder and a decoder component, and they comprise generalized stackable blocks, making it feasible to experiment with deeper models. Both the inputs and outputs are translated to an embedding space, but for the lack of convolutional or recurrent operations, the sequential or ordering information gets lost, which is compensated by using positional encoding. BERT, the next frontier in language modeling, outperformed the state-of-the-art models on 11 different tasks, including language understanding, question answering, multi-genre natural language inference. merely employing fine-tuning, thereby making task-specific architectural modifications redundant. Internally, BERT is a multi-layer bidirectional Transformer encoder [21], which comprises stacks of transformer blocks as shown in Fig. 1f. Although transformers are general sequence transduction models, BERT is more oriented towards being a language model. The BERT model is pretrained on 2 different tasks, namely, masked language modeling and next sentence prediction. After developing the pretrained BERT model, it can be conveniently adapted for most downstream NLP tasks just by finetuning the model using the new task-specific input-output pair, without any major modification. Ever since its introduction, BERT has become the defacto standard of transfer learning in NLP problems. The state-of-the-art performance of BERT is the result of a few contributory factors such as using stacks of transformer blocks that too with almost double attention heads, leveraging bidirectional information, and the use of segment embeddings.

**Figure 1**: Popular language models used in natural language processing. (a)word2vec being the simplest one, uses a shallow feed forward neural network to compute the word embeddings using long-linear models like continuous bag of words or skip-gram. LSTM models consider the context and are capable of producing better embeddings. The vanilla LSTM model has been augmented with improved regularization and expressivity in the different variants. (b) AWD-LSTM aims at better regularization by employing different kinds of dropouts in inputs, outputs, hidden layers, weights, and embeddings. Furthermore, a modified averaged stochastic gradient descent is employed during training. (c) mLSTM on the other hand attempts at increasing the expressivity of LSTM, by learning different recurrent transformations for different inputs. (d) ELMo improves such models further by leveraging bidirectional LSTM layers which enables ELMo to produce context-sensitive embeddings for words, unlike word2vec. (e) Transformers dispense recurrent layers and use attention instead. This not only makes the computations faster but also makes the process parallelizable, which resulted in a huge leap in NLP over the time-consuming RNN model training. (f) BERT improves the state of the art even further by employing stacks of bidirectional transformers. This deeper architecture accompanied by the availability of large-scale text corpora makes BERT the de facto standard in language modeling.

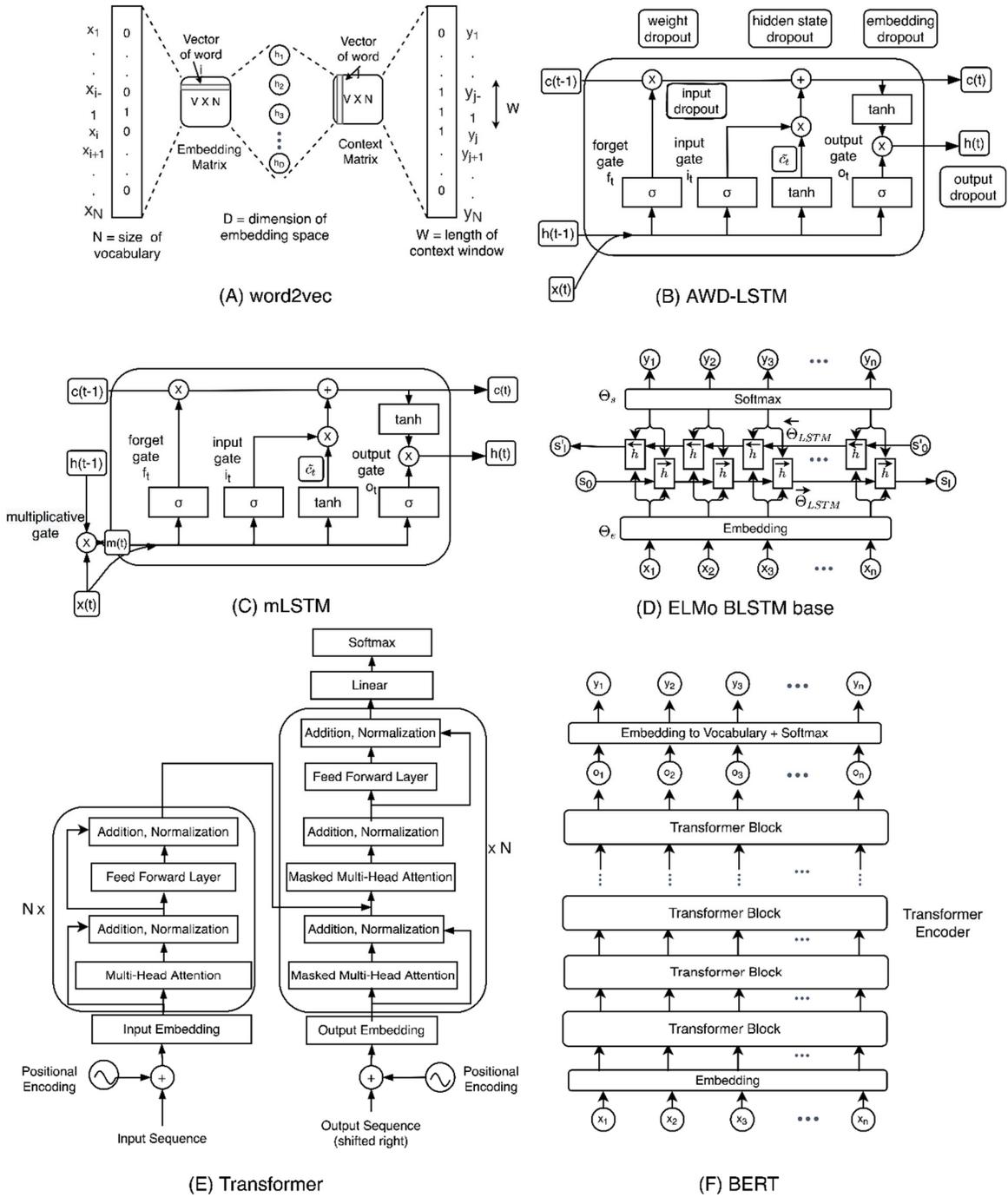

## 3 Protein databases facilitating language modeling

The primary goal of language model pretraining is to train it with a huge volume of data so that the model can learn diversified and distinct patterns in the language. Although managing such a high volume of data apparently seems challenging, the pretraining pipeline is simplified as it doesn't require any labeling of the data. This makes the task easier for NLP as we have huge volumes of unlabeled textual information sources like Wikipedia [22], which can be used for language model pretraining. Similarly, the combination of the diligent efforts of biologists over many decades with next-generation sequencing technologies has resulted in billions of protein sequence data. Thus, this opens up a huge scope for employing language models in protein sequence analysis.

Protein sequence databases, SwissProt [23], Pfam [24], and UniRef [25] are among the most notable ones. SwissProt is a manually curated protein sequence database, which provides high-quality annotations of 565,254 proteins. UniRef (UniProt Reference Clusters) provides clustering of sequences from the UniProtKB (UniProt Knowledgebase) and selected UniParc (UniProt Archive) records, i.e., a total of 216 million proteins, to obtain complete coverage as well as removing redundancy. The Pfam database is one of the most widely used resources to analyze protein families and domains, having a huge collection of 47 million proteins in 19,179 families.

Metagenome databases turned out to be very useful for training models due to their large size. BFD [26, 27] is a metagenome database containing around 2122 million protein sequences. BFD is the largest protein database at the time of writing, even eight times larger than the previous largest merged database [28]. MGnify is another metagenome resource, which contains over 1.1. billion sequences [81].

When training deep language models, the volume of the dataset plays an important role. For example, Heinzinger et al. [2] found that training an ELMo model with the SwissProt dataset resulted in less useful models, compared to training the same model with a larger UniRef50 dataset. For instance, when training the model on UniRef50 dataset a significant improvement was observed in downstream tasks such as secondary structure prediction, Localization prediction. Elnaggar et al. [13] similarly investigated the impact of database size on performance using three datasets, namely UniRef100, UniRef50, and BFD. It was seen that UniRef databases, particularly UniRef50 was sufficient and using BFD, which is 10 times larger than UniRef50, resulted in an almost minor improvement in classification accuracy and membrane protein prediction. But the outcome of using UniRef50 and BFD seemed to be task-dependent. For secondary structure prediction, UniRef50 performed on par with an existing method, NetSurfP-2.0 [83]. On the other hand, UniRef50 declined the model's performance by about 2% on tasks such as subcellular localization prediction, membrane protein prediction when a larger ProtT5-XXL model was trained in comparison with BFD.

# 4 Adapting Language Models for protein sequences

In this section, we highlight some of the major attempts in adapting language models for protein sequences to perform prediction tasks.

## 4.1 ProtVec (word2vec)

Asgari et al. [12], for the first time, applied the concept of word embedding in analyzing protein sequences. They used the skip-gram-based word2vec model to generate embeddings from 3-mers of amino acids. The word2vec model was trained on 546,790 manually annotated sequences from SwissProt. In order to tokenize the sequences, they took 3 shifted versions of the sequence and broke them into non-overlapping 3-mers. These 3-mers are thus treated as words and the broken sequences are treated as sentences, with which the word2vec model was trained with negative sampling. In addition to selecting the suitable contexts for a word, negative sampling randomly picks some contexts which are not related to that word, thus it enables the embeddings of similar words proximate and dissimilar words distant. The trained model, named as ProtVec, can embed 3-mers in 100-dimensional space. The learned protein space is also consistent in terms of the distribution of various biophysical and biochemical properties such as volume, polarity, hydrophobicity, etc.

The model was applied in two tasks, protein family classification, and disordered protein region prediction. They used the computed 100 dimensional 3-mer vectors, averaged over the entire sequence as the input to an SVM classifier. The family classification, applied on 324,018 protein sequences from SwissProt, spanning over 7,027 protein families, showed an average accuracy of 93%, surpassing existing methods. The accuracy of disordered protein prediction were 99.8% and 100% on the two datasets they used. ProtVec was later extended to ProtVecX [32], which can operate on variable-length segments of protein sequences. The extension of word2vec, doc2vec has also been applied for protein sequence analysis tasks, e.g., protein localization prediction [33].

Since this foundational work of Asgari et al. [12], word2vec were used in diverse prediction tasks, including protein-protein interaction binding sites [34], compound-protein interaction [35], protein Glycation sites [36], generalized protein classification [37]. Word2vec embeddings were used as input for conventional machine learning models such as SVM, KNN, Random Forest [38, 39, 40] and also for deep learning models, e.g. CNN [36], RNN [41], and Transformers [35].

In some works, FastText skip-gram model [42], which represents each word as a bag of character n-gram, was also employed to generate embeddings from protein sequences, which was later used for different types of analysis [38, 40, 43]. Islam et al. [37] proposed m-NGSG, which modifies the k-skip-bi-gram model by employing a combination of n-grams and skip grams and demonstrated consistent improvement on tasks including localization prediction, fluorescent protein prediction, antimicrobial peptide prediction. Ideally, the embeddings are expected to work well in many prediction tasks, but there is a report that a network specifically trained on a particular task (kinase-substrate phosphorylation prediction) showed a better performance than ProtVec [44]. Ibtehaz et al. [45] analyzed and found that the ProtVec embeddings hardly correspond with similarity scores of the k-mers, i.e. the vector similarity (cosine similarity) of the embeddings of the k-mers correlates little with the similarity score (alignment score) of the k-mers (Pearson correlation coefficient of 0.226). Ibtehaz et al. proposed the Align-gram model, which modified the Skip-gram model to make it more aligned with protein analysis, making vector similarity and k-mer similarity equivalent (Pearson correlation coefficient of 0.902).

## 4.2 UDSMProt (AWD-LSTM)

Strodthoff et al. attempted to devise a single, universal model architecture to solve diverse problems related to proteins. The proposed architecture UDSMProt [29] was based on the AWD-LSTM language model [18], which is internally a 3-layer LSTM network, with different dropout regularizations (input, embedding, weight, hidden state, and output layer dropout). The model was pretrained on SwissProt.

UDSMProt operates on protein sequence data tokenized to the amino acid level. The pretraining aimed towards predicting the next token for a given sequence of tokens, implicitly learning the structure and semantics of the language i.e., protein sequences. During various downstream prediction tasks, the embeddings obtained from the model were compiled through a Concat-Pooling layer and some dense layers were added on top which is trained in the process of finetuning. The UDSMProt pipeline was evaluated on three different tasks, enzyme class prediction, gene ontology prediction, and remote homology and fold detection. With mere finetuning on the problem datasets, the proposed method performs on par with state-of-the-art algorithms that were tailored to those specific tasks, even surpassing them in two tasks.

UDSMProt was later used in the USMPEP [63] pipeline. State-of-the-art result was obtained on MHC class I binding prediction, using just a generic model without any domain-specific heuristics.

## 4.3 UniRep (mLSTM)

Alley et al. [31] trained an mLSTM model with 1900 hidden units, UniRep, on around 24 million protein sequences in UniRef50. Despite being trained in an unsupervised manner, i.e., predicting the next amino acid from a sequence of amino acids, UniRep embeddings managed to create physicochemically meaningful clusters of amino acids and partition structurally similar protein sequences. The learned embeddings were able to predict protein secondary structure, the stability of natural and de novo designed proteins, and the quantitative function of molecularly diverse mutants. It was also shown that UniRep has the potential to enhance efficiency in protein engineering tasks, as demonstrated in predicting fluorescence in engineered proteins. The primary contribution of UniRep was extracting the fundamental protein features using unsupervised deep learning as fixed-length vectors which are both semantically rich and structurally, evolutionarily, and biophysically grounded.

In addition, UniRep embedding-based feature representation has demonstrated improved performance in other tasks, anticancer peptides prediction [64], assessing disease risk of protein mutation [65], localizing sub-Golgi proteins [66], and Peroxisomal proteins [52]. UniRep embeddings correlated with biological features important for protein expression in *B. subtilis* [67] and can also be used to analyze interaction patterns between virus and human proteins [68].

## 4.4 SeqVec (ELMo)

As mentioned in earlier sections, word2vec embeddings ignore context, which can be solved by using a complex language model like ELMo. Heinzinger et al. [2] proposed SeqVec, which is an ELMo model trained on protein sequences. The authors basically used the standard ELMo implementation. The two-layer ELMo model applied dropout and shared weights between forward and backward LSTMs to reduce overfitting. It was trained on the UniRef50 database. The SeqVec model can take a protein sequence and returns 3076 features for each residue in the sequence.

The embeddings generated from SeqVec were evaluated in four different tasks, namely, secondary structure prediction, disorder prediction, localization prediction, and membrane prediction and showed better performance than other sequence-based representations such as one-hot encoding and ProtVec. SeqVec runs much faster than evolutionary methods e.g. HHBlits [46] and the speed is not affected by the size of the database, thus is massively scalable.

With the release of SeqVec [2], ELMo has been promptly received as a welcomed addition to the bioinformatics analysis toolbox. Zeng et al. [47] used ELMo to learn a context-dependent embedding of amino acids for MHC I class ligand prediction. Litmann et al. [48] demonstrated that by simply using ELMo and BERT-based embeddings, it is possible to almost reach the state of the art in protein function prediction. Again, Villegas et al. [49] leveraged ELMo embedding as features for protein function prediction. SeqVec features was also applied to B-cell epitope prediction [50], and cross-species protein function prediction [51]. Moreover, Elmo embeddings were used as input for SVM [52] and Graph Neural Networks [53].

Apart from the standard ELMo architecture, general BLSTM networks have also been used in several other protein language modeling tasks. Bepler et al. [75] trained a multitask neural network to solve protein structural tasks, contact prediction and structural similarity prediction by training on protein structure information. In another work [76], the authors also experimented with introducing a two-stage feedback mechanism where they trained a BLSTM on protein sequences and contact map information and a proposed 'soft symmetric alignment'. Primarily, the encoder generates embeddings from the amino acid sequence. Later the embeddings are used to predict contact maps and compute L1 distance between pairs of proteins by the proposed soft symmetric alignment. These error terms are fed as feedback signals to the language model, thus making the embeddings more biologically driven. DeepBLAST [77] on the other hand, obtained alignments from embeddings learned from the protein language model in [75] and integrates them into an end-to-end differentiable alignment framework.

## 4.5 ESM-1b (Transformer)

Rives et al. [69] trained transformer models on 250 million protein sequences from UniParc [3]. Initially, transformer models with 100M parameters were trained and a systematic hyperparameter optimization was performed. After finalizing the suitable hyperparameter set, the model was scaled to 33 layers, having around 650M parameters. The trained ESM-1b transformer managed to learn the biochemical properties of the amino acids. The output embeddings allowed to cluster the residues in several groups which are consistent with the hydrophobic, polar, and aromatic nature of amino acids. Furthermore, the molecular weight and charge information was also reflected across the amino acids. Moreover, the different biological variations are encoded in the representation space. Specifically, the embeddings without any explicit information managed to cluster the orthologous genes together. Furthermore, the learned embeddings are suitable to be used as feature representations for various downstream tasks. The authors demonstrated applications of the trained model in remote homology prediction, secondary structure prediction, and contact prediction. ESM-1b embeddings were used in protein function prediction [70], effects of mutations on protein function [71], contact map prediction [72], protein fitness prediction [73], and Lectin-Glycan Binding Prediction [74].

A recent work based on transformer language modeling, MSA transformer [84] used multiple sequence alignment as inputs to a transformer and significantly improves the performance over ESM-1b in unsupervised contact prediction, increasing top-L long-range contact precision by 15 points. MSA transformer also outperforms NetSurfP-2.0 in secondary structure prediction by increasing in Q8 accuracy by 2%.

## 4.6 ProtTrans (BERT)

Since its introduction, BERT has become the defacto standard model of solving NLP problems. Elnaggar et al. [13] ported BERT for protein sequence analysis. They trained two auto-regressive models (Transformer-XL, XLNet) and four auto-encoder models (BERT, Albert, Electra, T5) on data from UniRef and BFD, using roughly 2,122 million protein sequences. The authors followed the standard implementations of the transformer models and trained different instances on different datasets. Training such networks on the astounding amount of data required the assistance of HPC (High-Performance Computing), using 5616 GPUs and TPU Pod up to 1024 cores.

The embeddings captured various biophysical properties of the amino acids, structure classes of proteins, domains of life and viruses, and protein functions in conserved motifs. The embeddings were also evaluated on per-residue (protein secondary structure prediction) and per-protein (cellular localization and membrane protein classification) levels. No task-specific modifications were performed, rather, the models were used as static feature extractors, by extracting embeddings derived from the hidden state of the last attention layer. From the experiments, it was observed that both for localization and secondary structure prediction fine-tuning improved performance. Impressively, embeddings from their trained ProtT5 model, for the first time, surpassed the state-of-the-art methods in the secondary structure prediction task, without using any evolutionary information. The authors assessed the impact of database size. They observed that models trained on UniRef50 were enough and adding the huge amount of data from BFD hardly presented noticeable improvements.

Despite BERT being just recently adopted for proteins, it has rapidly gained popularity. Hiranuma et al. [54] used ProtBERT embeddings along with several structural features to guide and improve protein structure refinement. Litmann et al. [48] investigated the effectiveness of BERT embeddings in Gene Ontology prediction. Charoenkwan et al. [55] used BERT embeddings to predict amino acid sequences of peptides that taste bitter without using any structural information and greatly outperformed the existing works. Filipavicius et al. [56] pretrained a RoBERTa model on a mixture of binding and random protein pairs and achieved enhanced downstream protein classification performance for tasks such as homology prediction, localization prediction, protein-protein interaction prediction as compared to the ESM-1b[69] transformer. Application of BERT embeddings was able to improve several peptide prediction tasks [57, 58, 59]. BERT embeddings were also effective as input representations for clustering algorithms [60] and graph neural networks [61, 62] for tasks such as clustering protein functional families [60], predicting effects of mutation [61], and protein-protein interaction site prediction [62].

Vig et al. [78] analyzed underlying learned information of protein transformer models, utilizing attention mechanisms. They analyzed and experimented with transformer models from TAPE [30] and ProtTrans [13], with a specific focus on the attention mechanism. Their analysis revealed that attention can capture high-level structured properties of proteins, namely, amino acids that were nearby in the 3D structure, despite being further apart in the 1D sequence. Furthermore, it was found that attention reflects binding sites, amino acid contact maps, and amino acid substitution matrices.

**Table 1: The list of reviewed methods in this article.**

| Embedding | Problem | Model | Additional Features | Dataset | Ref | Source Code/Server |
|---|---|---|---|---|---|---|
| ProtVec (word2vec) | Protein-protein interaction (PPI) binding sites prediction | CNN RNN fusion | HSP, PSSM, ECO, RSA, RAA, disorder, hydropathy, physicochemcial properties | Recent publications | [34] | https://github.com/lucian-ilie/DELPHI  Server: https://delphi.csd.uwo.ca |
| | Compound–protein interaction prediction | Transformer | Atomic properties | Human dataset, Caenorhabditis elegans dataset, BindingDB dataset | [35] | https://github.com/lifanchen-simm/transformerCPI |
| | Protein glycation sites prediction | LSTM | - | 3 Surveyed Datasets | [36] | Server: http://watson.ecs.baylor.edu/ngsg |
| | Protein classification | LR | - | Subchlo, osFP, iAMP-2L, Cypred and PredSTP, TumorHPD 1 and 2, HemoPI 1 and 2, IGPred and PVPred | [37] | https://bitbucket.org/sm_islam/mngsg/src/master/ |
| | Transporter substrate specificities identification | SVM | - | Proteins involved in trans-porting ion/molecules, collected from UniProt (release 2018_10). Dataset available in : http://bio216.bioinfo.yzu.edu.tw/fasttrans/ | [38] | Server: http://bio216.bioinfo.yzu.edu.tw/fasttrans |
| | Nuclear localization signal identification | Multivariate Analysis | physicochemcial properties, disorder, PSSM | NLSdb 2003, NLSdb 2017, SeqNLS | [39] | Server: http://www.csbio.sjtu.edu.cn/bioinf/INSP/ |
| | Tumor necrosis factors idetifcation | SVM | - | 106 protein from tumor necrosis factor family and 1023 sequences from other major cytokine families were collected from UniProt (release 2019_05) | [40] | https://github.com/khucnam/TNFPred |
| | MHC binding prediction | GRU | MHC Allele embedding | IEDB and rececnt publications | [41] | https://github.com/cmb-chula/MHCSeqNet |

| | Nucleic acid-binding protein identification | NN | RNA sequence embedding | RNAcompete dataset, PBM dataset and recent publications | [43] | https://github.com/syang11/ProbeRating |
|---|---|---|---|---|---|---|
| UDSMProt (AWD-LSTM) | MHC binding prediction | LSTM | - | IEDB, HPV | [63] | https://github.com/nstrodt/USMPep |
| UniRep. (mLSTM) | Anticancer peptide prediction | KNN, LDA, SVM, RF, LGBM, NB | Pretrained SSA embedding | AntiCP 2.0 datasets | [64] | https://github.com/zhibinlv/iACP-DRLF |
| | Disease risk prediction | MLP | Hydrophilic properties | BRCA1, PTEN | [65] | https://github.com/xzenglab/BertVS |
| | Sub-Golgi localization identification | SVM | - | Recent publications | [66] | https://github.com/zhibinlv/isGP-DRLF  Server: http://isgp-drlf.aibiochem.ne |
| | Peroxisomal proteins localisation prediction | SVM | SeqVec embedding | Protein sequences for peroxisomal membrane and matrix proteins collected from UniprotKB/SwissProt database. Dataset available in : https://github.com/MarcoAnteghini/In-Pero/tree/master/Dataset | [52] | https://github.com/MarcoAnteghini/In-Pero |
| SeqVec (ELMo) | MHC class I ligand prediction | Residual Network | One hot encoding, BLOSUM50 | Recent publications | [47] | https://github.com/gifford-lab/DeepLigand |
| | Protein function prediction | Modified KNN | BERT embedding | CAFA3 | [48] | https://github.com/Rostlab/goPredSim  Server: https://embed.protein.properties |
| | Protein function prediction | KNN, LR, MLP, CNN, GCN | One hot encoding, k-mer, DeepFold features, Contact map | CAFA3 | [49] | https://github.com/stamakro/GCN-for-Structure-and-Function |
| | linear B-cell epitope prediction | NN | Amino acid embedding | IEDB Linear Epitope Dataset | [50] | https://github.com/mcollatz/EpiDope |
| | Protein function prediction | LR | - | SwissProt, cross-species datasets | [51] | |

| | Peroxisomal proteins localisation prediction | SVM | UniRep embedding | Protein sequences for peroxisomal membrane and matrix proteins collected from UniprotKB/SwissProt database. Dataset available in : https://github.com/MarcoAnteghini/In-Pero/tree/master/Dataset | [52] | https://github.com/MarcoAnteghini/In-Pero |
|---|---|---|---|---|---|---|
| | Cofactor specificity of Rossmann-fold protien prediction | GCN | - | ECOD and literature datasets | [53] | https://github.com/labstructbioinf/rossmann-toolbox<br><br>Server: https://lbs.cent.uw.edu.pl/rossmann-toolbox |
| ESM-1b (Transformer) | Protein function prediction | GAT | inter-residue contact graphs | PDB-cdhit | [70] | |
| | Effect of mutation prediction | Transformer | - | 41 deep mutational scans | [71] | |
| | Contact map prediction | CNN | One hot encoding, SS3, SS8, ASA< HSE, protein backbone torsion angles | ProteinNet, CASP14-FM, SPOT-2018 | [72] | https://github.com/jas-preet/SPOT-Contact-Single<br><br>Server: https://sparks-lab.org/server/spot-contact-single/ |
| | Protein fitness prediction | Ridge Regression | One hot encoding, physicochemical representaion | 19 labelled mutagenesis datasets | [73] | |
| | Lectin-Glycan binding prediction | MLP | SweetNet features | Dataset was curated from 3,228 glycan arrays from the Consortium for Functional Glycomics database and 100 glycan arrays from the Carbohydrate Microarray Facility of Imperial College London | [74] | https://github.com/BojarLab/LectinOracle |
| ProtTrans (BERT) | Protein structure refinement | CNN | distance maps, amino acid identities and properties, backbone angles, residue angular orientations, Rosetta energy terms, secondary structure information, MSA information | PISCES | [54] | https://github.com/hiranumn/DeepAccNet |

| Protein function prediction | Modified KNN | ELMo embedding | CAFA3 | [48] | https://github.com/Rostlab/goPredSim<br><br>Server: https://embed.protein.properties |
|---|---|---|---|---|---|
| Peptide binding site identification | Transformer | - | peptide complex dataset | [57] | |
| Signal peptide prediction | CRF | - | Extended previously published dataset with newly available sequences from UniProt, Prosite and TOPDB | [58] | Server: https://services.healthtech.dtu.dk/service.php?SignalP-6.0 |
| MHC-peptide class II interaction prediction | Transformer | - | IEDB and recent publications | [59] | https://github.com/s6juncheng/BERTMHC<br><br>Server: https://bertmhc.privacy.nlehd.de |
| Functional family clustering | DBSCAN | - | CATH | [60] | https://github.com/Rostlab/FunFamsClustering |
| Effect of mutation prediction | LGBM | ProteiSolver features | ProTherm, SKEMPI | [61] | Server: http://elaspic.kimlab.org |
| Protein-protein interaction (PPI) binding sites prediction | GCN | - | Recent publications | [62] | https://github.com/Sazan-Mahbub/EGRET |

For each method, we listed the problem solved, machine learning model, additional features, dataset, and software availability.
Abbreviations : CNN = Convolutional Neural Network, RNN = Recurrent Neural Network, LSTM = Long Short-Term Memory, LR = Logistic Regression, SVM = Support Vector Machine, GRU = Gated Recurrent Unit, NN = Neural Network, KNN = K-Nearest Neighbors, MLP = Multi-Layer Perceptrons, GCN = Graph Convolutional Network, DBSCAN = Density-Based Spatial Clustering of Applications with Noise, CRF = Conditional Random Field, LGBM = Light Gradient Boosting Machine, LDA = Latent Dirichlet Allocation, RF = Random Forest, NB = Naive Bayes, GAT = Graph Attention

# 5 Conclusion

Various successful adaptations of the language models in bioinformatics have greatly benefitted the analysis of protein sequences and various kinds of predictions. In addition to using embedding learning on specific datasets and tasks [79,80], pretraining language models on millions of protein sequences can dramatically improve the performance of downstream tasks.

We have reviewed a total of 33 methods (as summarized in Table 1) that rely on sequence embeddings as input, moving away from the traditional bioinformatics pipeline of computing PSSM or HMM profiles. Among the various types of embeddings, ProtVec and ProtTrans have been used most frequently (in 9 and 8 methods respectively). For various problems, such as secondary structure prediction, protein-protein interaction site prediction, previous state-of-the-art performance have already been surpassed merely using embeddings learned in an unsupervised manner. We expect to observe that embedding techniques and pre-trained embeddings will be applied in many other tasks and make substantial improvements in the field.

**Acknowledgement**
This work was partly supported by the National Institutes of Health (R01GM133840, R01GM123055, and 3R01GM133840-02S1) and the National Science Foundation (CMMI1825941, MCB1925643, and DBI2003635).